\documentclass[journal=aamick,manuscript=article]{achemso}

\usepackage{color}
\usepackage{graphicx}
\usepackage{dcolumn}
\usepackage{bm}
\usepackage{amsmath}
\usepackage{wrapfig}
\usepackage{graphicx}
\usepackage{xr}
\usepackage{gensymb}

\setkeys{acs}{maxauthors=0}
\setkeys{acs}{chaptertitle=true}
\setkeys{acs}{articletitle=true}

\usepackage{xr}
\makeatletter
\newcommand*{\addFileDependency}[1]{
  \typeout{(#1)}
  \@addtofilelist{#1}
  \IfFileExists{#1}{}{\typeout{No file #1.}}
}
\makeatother

\title{Cold Seeded Epitaxy and Flexomagnetism in Smooth GdAuGe Membranes Exfoliated from graphene/Ge(111)}

\author{Zachary LaDuca}
\author{Tamalika Samanta}
\author{Nicholas Hagopian}
\author{Taehwan Jung}
\author{Katherine Su}
\affiliation{Materials Science and Engineering, University of Wisconsin-Madison, Madison, WI 53706, United States of America}
\author{Konrad Genser}
\author{Karin M. Rabe}
\affiliation{Physics and Astronomy, Rutgers University, NJ 08854, United States of America}
\author{Paul M. Voyles}
\author{Michael S. Arnold}
\author{Jason K. Kawasaki} \email{jkawasaki@wisc.edu}
\affiliation{Materials Science and Engineering, University of Wisconsin-Madison, Madison, WI 53706, United States of America}

\date{\today}

\begin{document}

\begin{abstract}

Remote and van der Waals epitaxy are promising approaches for synthesizing single crystalline membranes for flexible electronics and discovery of new properties via extreme strain; however, a fundamental challenge is that most materials do not wet the graphene surface. We develop a cold seed approach for synthesizing smooth intermetallic films on graphene that can be exfoliated to form few nanometer thick single crystalline membranes. Our seeded GdAuGe films have narrow x-ray rocking curve widths of 9-24 arc seconds, which is two orders of magnitude lower than their counterparts grown by typical high temperature methods, and have atomically sharp interfaces observed by transmission electron microscopy. Upon exfoliation and rippling, strain gradients in GdAuGe membranes induce an antiferromagnetic to ferri/ferromagnetic transition. Our smooth, ultrathin membranes provide a clean platform for discovering new flexomagnetic effects in quantum materials.

\end{abstract}

\maketitle

\textbf{keywords:} graphene, epitaxy, membrane, Heusler, strain

Remote \cite{kim2017remote, kong2018polarity} and van der Waals epitaxy \cite{koma1992van, shi2012van, kim2014principle, du2022controlling} on graphene-covered surfaces are promising strategies for synthesizing single crystalline membranes. Here, the weak van der Waals interactions between the film and graphene enable etch-free membrane exfoliation. These membranes enable applications for flexible electronics \cite{kim2017remote} and discovery of new properties induced by extreme strain \cite{hong2020extreme, harbola2021strain, du2021epitaxy, dong2019super}. For example, flexomagnetism is magnetic ordering induced by strain gradients \cite{lukashev2010flexomagnetic}. While there has been some success in producing strain gradients in epitaxial thin films via partial relaxation \cite{zhang2021strain, makushko2022flexomagnetism}, greater opportunities for control exist for bending of ultrathin single crystalline membranes \cite{du2023strain}. This includes experimental demonstration of an antiferromagnetic to ferromagnetic transition in rippled GdPtSb membranes \cite{du2021epitaxy}, predicted cycloidal spin textures induced by bending in CrI$_3$ \cite{edstrom2022curved}, and predicted bending-induced skyrmions via control of Dzyaloshinskii-Moriya Interaction (DMI) in CrBr$_2$ and MnSe$_2$ \cite{ga2022dzyaloshinskii}, since strain gradients and bending break inversion symmetry.

There are two significant challenges for epitaxy on graphene. First, since graphene has very low surface energy, most materials do not wet the graphene surface and instead grow with a disconnected island morphology. This includes GaN islands on graphene/SiC \cite{qiao2021graphene}, oxide islands on graphene/SrTiO$_3$ \cite{chang2023remote}, III-V islands \cite{alaskar2014towards, zulqurnain2022defect, manzo2022pinhole} or nanowires \cite{hong2012van} on graphene, and metal islands on graphene \cite{liu2013metals, laduca2023control}. The wetting challenge limits the ability to access the largest strains that are only possible in the ultrathin limit: whereas $8\%$ strain has been reported for few unit cell thick (La,Ca)MnO$_3$ (LCMO) membranes, LCMO with thickness greater than 20 nm could only undergo $2\%$ strain before cracking \cite{hong2020extreme}. Second, the sticking coefficients on graphene can be highly temperature and element dependent, which poses challenges for controlling the stoichiometry of alloys and compounds that are not adsorption-controlled \cite{laduca2023control}.

Here we demonstrate a cold seeded epitaxy approach to solve the dewetting and sticking coefficient challenges. Using molecular beam epitaxy (MBE) of GdAuGe on graphene/Ge and graphene/SiC, we demonstrate few nanometer thick films on graphene with a smooth step and terrace morphology and a 9-24 arc sec x-ray rocking curve width that is two orders of magnitude sharper than their counterparts grown at high temperature. The cold seed suppresses surface diffusion during nucleation, while subsequent annealing and continued growth at elevated temperatures recrystallizes the seed. The cold seed also provides a template with a well defined stoichiometry, since sticking coefficients on graphene at room temperature are much closer to unity. The GdAuGe films can be exfoliated to produce continuous membranes with minimal tears that are supported on a polymer handle. We demonstrate that strain gradients in rippled GdAuGe membranes induce an antiferromagnetic to ferromagnetic transition, similar to that observed for GdPtSb membranes \cite{du2021epitaxy}. We anticipate that this strategy is generally applicable to epitaxy of other thin film systems on graphene.

\begin{figure}[h]
    \centering
    \includegraphics[width=0.45\textwidth]{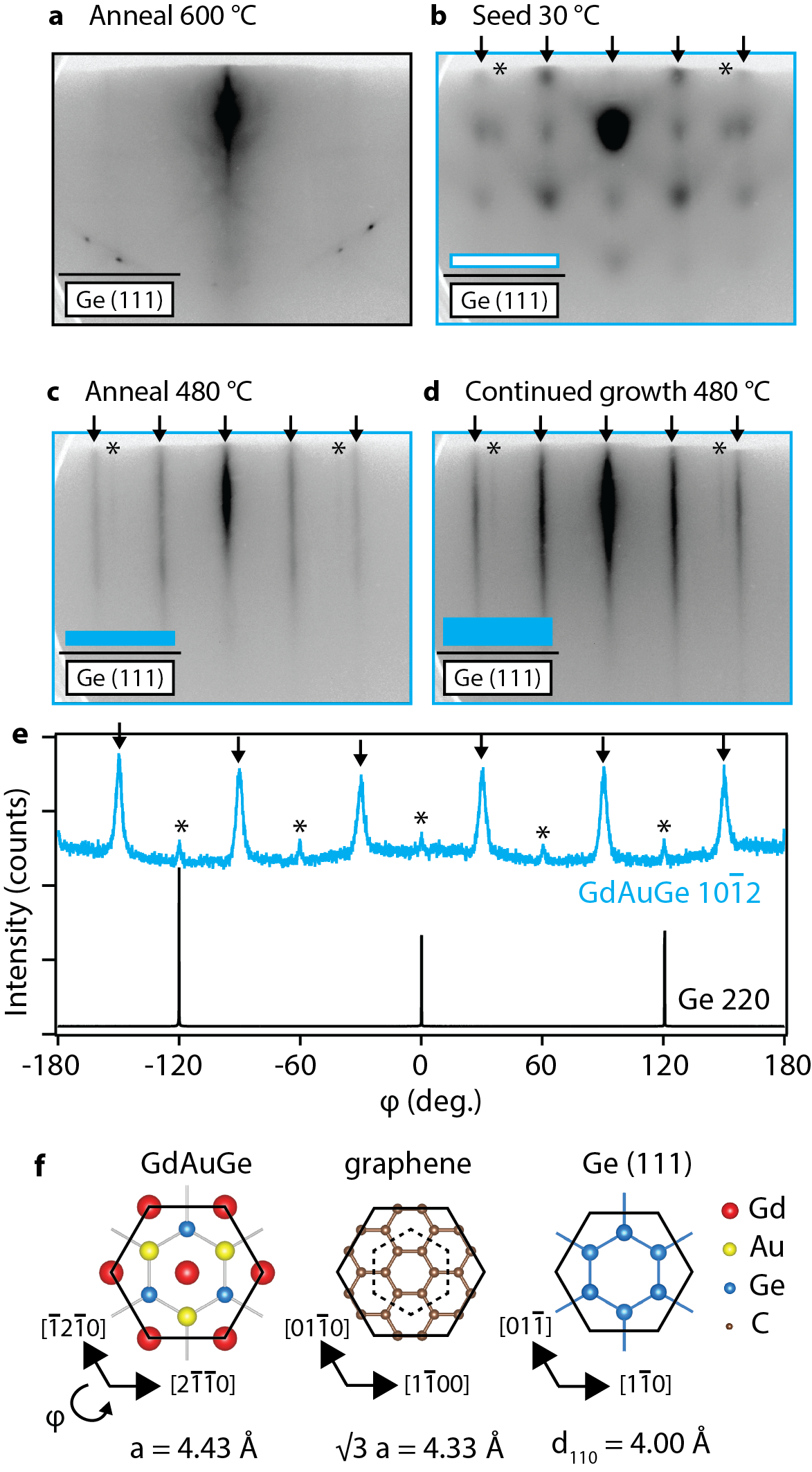}
    \caption{\textbf{Cold seeded epitaxy of GdAuGe on graphene.} (a) RHEED pattern of graphene on Ge (111), after annealing at $600 \degree$C to remove surface adsorbates. All RHEED patterns are recorded along a Ge $\langle 110 \rangle$ zone axis at a beam energy of 15 kV.  (b) 5 nm thick seed of GdAuGe grown at $30\degree$C. (c) Seed after recrystallizing by annealing at $480\degree$C. (d) Continued growth of GdAuGe to a total thickness of 18 nm. Arrows mark the primary GdAuGe and graphene domain reflections. Stars mark the 30 degree rotated orientation. (e) X-ray diffraction azimuthal $\phi$ scan showing the in-plane epitaxial alignment. (f) Crystal structures of GdAuGe, graphene, and Ge showing the primary epitaxial alignment. For graphene, the primitive unit cell is marked by dotted lines and a $(\sqrt{3}\times \sqrt{3})R30 \degree$ supercell is marked by solid lines.}
    \label{seed}
\end{figure}

Figure \ref{seed} illustrates the key steps in our seeded growth approach. We start from graphene that is grown by chemical vapor deposition on Ge (111), following Ref \cite{kiraly2015electronic}. After annealing the graphene in ultrahigh vacuum ($< 2 \times 10^{-10}$ Torr) at $600\degree$ C for one hour, the reflection high energy electron diffraction (RHEED) pattern along the Ge $[1 \bar{1} 0]$ azimuth exhibits a sharp ring of spots (Fig. \ref{seed}a), indicating a nearly atomically smooth and clean surface. Previous low energy electron diffraction studies show that epitaxial graphene on Ge (111) has two domain orientations that are rotated 30 degrees from one another: a primary domain with graphene $[1 \bar{1} 0 0] \parallel$ Ge $[1 \bar{1} 0]$ and a secondary domain with graphene $[1 \bar{2} 1 0] \parallel$  Ge $[1 \bar{1} 0]$ \cite{kiraly2015electronic}. 

To suppress the dewetting and island formation that are typical for high temperature growth of metals on graphene surfaces \cite{laduca2023control}, we initiate growth of a 5 nm thick GdAuGe seed at room temperature ($30\degree$C). Here, equiatomic fluxes of Gd, Au, and Ge were supplied by thermal effusion cells at rates of $1.2 \times 10^{13}$ atoms/cm$^2$s, respectively, as measured by quartz crystal microbalance and calibrated by Rutherford backscattering spectrometry (RBS) on test samples. The corresponding RHEED pattern indicates an epitaxial seed with a semi-rough surface: the RHEED spots are connected by extended streaks with well-defined lattice spacings (Fig. \ref{seed}b). Subsequent annealing of the seed layer at $480 \degree$C recrystallizes it and smooths the surface, as evidenced by the sharp and streaky RHEED pattern in Fig. \ref{seed}(c).

The smooth surface of the annealed seed layer serves as a template for continued GdAuGe growth at elevated temperatures. Fig. \ref{seed}(d) shows the RHEED pattern of a GdAuGe film regrown at $480\degree$ C to a total thickness of 16 nm GdAuGe. The bright, sharp, and extended streaks indicate a flat epitaxial film. The final GdAuGe surface exhibits two sets of streaks: a primary set with with spacing $\Delta Q$ (arrows), and a secondary set with spacing $\sqrt{3} \cdot \Delta Q$ (asterisks), indicative of two GdAuGe domains that are rotated 30 degrees from one another. An azimuthal $\phi$ scan by x-ray diffraction confirms two GdAuGe orientations rotated around the $c$ axis by 30 degrees (Fig. \ref{seed}(e)).

An interesting question is whether GdAuGe films on graphene/Ge are epitaxial to the graphene or to the Ge substrate, e.g., via ``remote'' substrate interactions that permeate through graphene \cite{kim2017remote}. Here we find a direct correspondence between two graphene orientations rotated by 30 degrees, and two GdAuGe orientations rotated by 30 degrees. This implies that the GdAuGe is epitaxial to graphene, since our control experiment of GdAuGe epitaxy directly on Ge (111) produces a single orientation of GdAuGe. The lattice mismatch of $2.3\%$ between GdAuGe and a $(\sqrt{3}\times \sqrt{3})R30 \degree$ supercell of graphene is smaller than the $9.7\%$ mismatch between GdAuGe and Ge (111) (Fig. \ref{seed}(f)), which also suggests a preferred epitaxial relationship to graphene. As an additional test, we grew GdAuGe on graphene/Ge(110), where the Ge (110) surface has rectangular symmetry compared to the hexagonal symmetry of GdAuGe. We find that GdAuGe again grows (0001)-oriented, with the same in-plane orientation with respect to graphene: GdAuGe $[2 \bar{1} \bar{1}0] \parallel$ graphene $[1 \bar{1} 0 0 ]$ (Supplemental Figure 1).

\begin{figure*}[ht]
    \centering
    \includegraphics[width=1\textwidth]{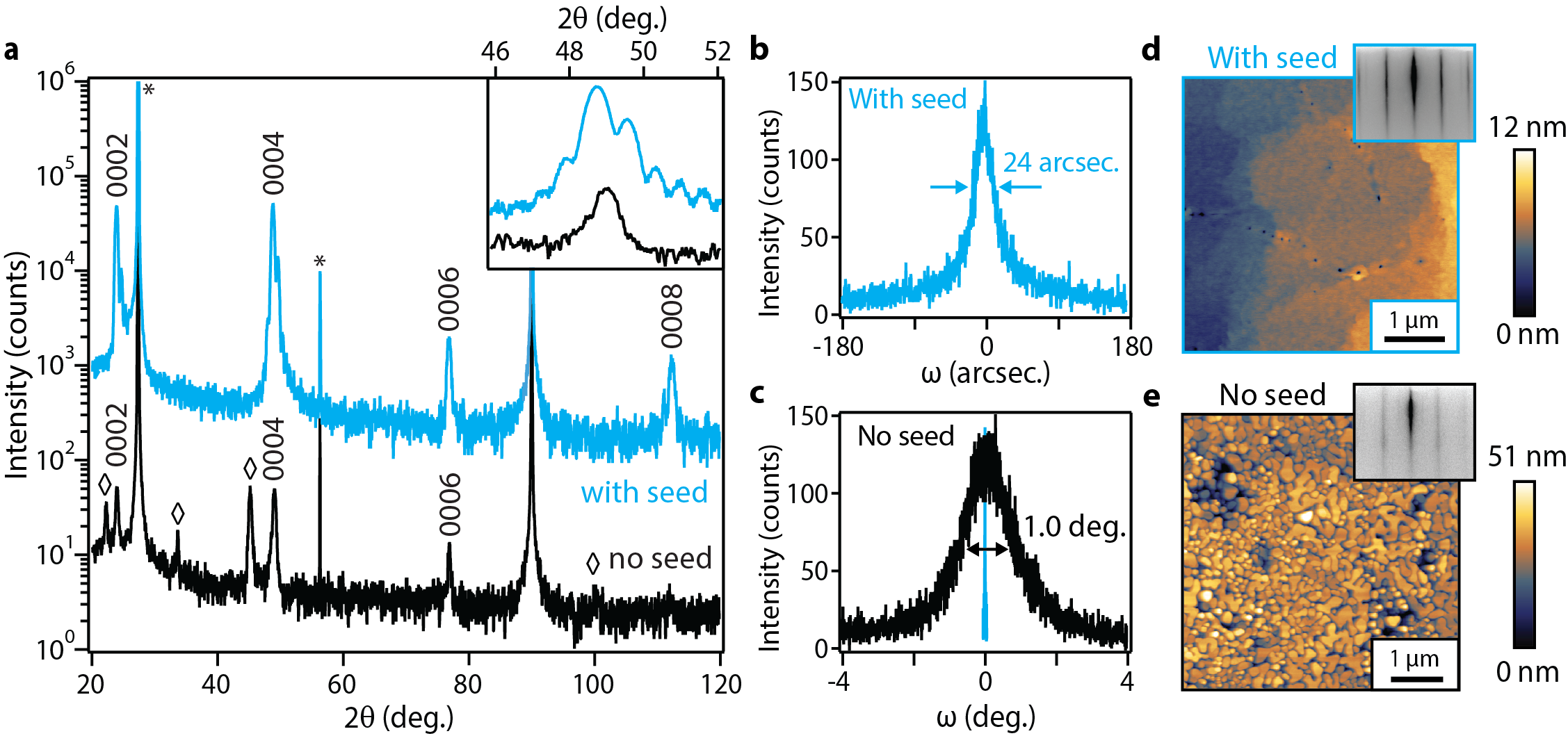}
    \caption{\textbf{Comparison of low temperature seeded growth versus direct high temperature growth of GdAuGe on graphene/Ge (111).} (a) X-ray diffraction scan (Cu $K\alpha$) of a film grown using the low temperature seed approach (blue) versus a film grown directly at high temperature (black). Insert shows the $0004$ reflection. (b,c) Rocking curves of $0004$ reflection for the the seeded and unseeded samples. (d,e) Atomic force microscopy showing a step and terrace morphology for the seeded sample and an island morphology for the no seed sample. Inserts show the RHEED patterns at the end of GdAuGe growth.}
    \label{Ge}
\end{figure*}

Figure \ref{Ge} compares the structure and morphology of a 16 nm GdAuGe film grown with a cold seeded growth approach on graphene/Ge(111) to that of an 18 nm film grown directly on graphene/Ge(111) at $540 \degree$C. For the seeded sample, the $2\theta - \omega$ x-ray diffraction scan displays bright GdAuGe $000l$ reflections, including the higher order $0006$ and $0008$ Bragg reflections. The sharp Kiessig fringes around the GdAuGe $0004$ reflection (insert) indicate a smooth surface and interface morphology (Fig. \ref{Ge}(a), blue trace). These x-ray fringes are consistent with the narrow x-ray rocking curve width of 24 arc seconds that indicates a low mosaicity (Fig. \ref{Ge}(b)). This rocking curve width is significantly narrower than the 245 arc sec reported for GaAs on graphene/SiO$_2$ \cite{alaskar2014towards}, 300 arc sec for GaN on graphene/GaN \cite{badokas2021remote}, 115 arc sec for SrTiO$_3$ on graphene/SrTiO$_3$ \cite{yoon2022free}, 180 arc sec for VO$_2$ on graphene \cite{guo2019reconfigurable}, 770 arc sec for ZnO on graphene \cite{wang2022improved}, and 1.2 degrees (4320 arc sec) for BaTiO$_3$ on graphene/Ge (110) \cite{dai2022highly}. We observe a flat atomic step and terrace morphology by atomic force microscopy (AFM, Fig. \ref{Ge}(d)). In the AFM scan we also observe an arc of circular openings with depth of a few nanometers into the sample. We attribute these depressions to defects induced by a graphene grain boundary or wrinkle.

In contrast, the sample without seed layer displays weaker $000l$ reflections, indicative of poorer crystallinity, no Kiessig fringes, and XRD reflections from secondary phases (diamond symbols). We attribute the secondary phases to the fact that the sticking coefficients of metals on graphene deviate significantly from unity at elevated temperatures, as observed previously for other metals on graphene \cite{laduca2023control}, and confirmed for GdAuGe on graphene by RBS (Supplemental Fig. 2). AFM of the non seeded sample displays a disconnected island morphology (Fig. \ref{Ge}e), consistent with large rocking curve width of 1 degree that indicates large mosaicity (Fig. \ref{Ge}c).

\begin{figure*}[ht]
    \centering
    \includegraphics[width=1\textwidth]{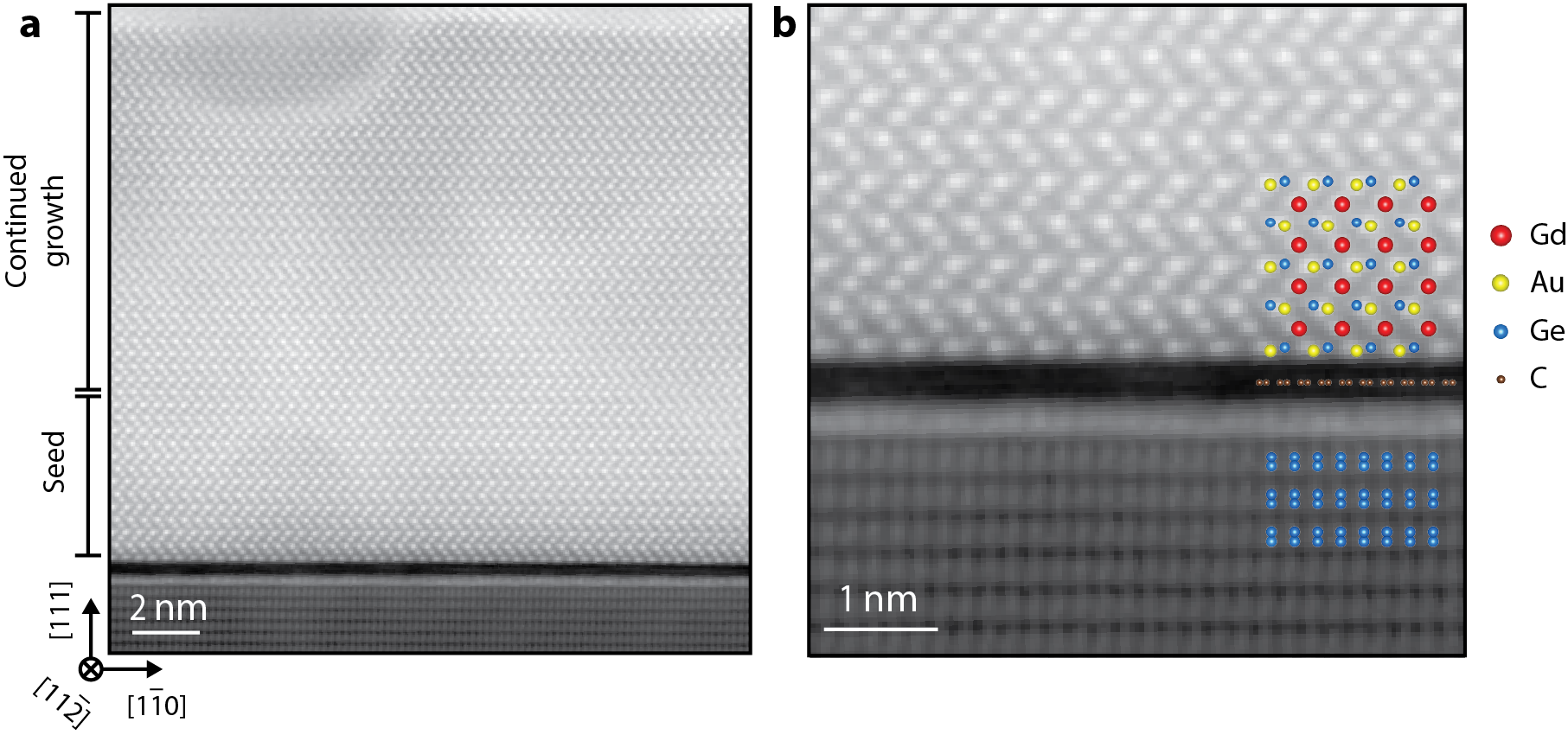}
    \caption{High-angle annular dark-field scanning transmission electron microscopy (HAADF-STEM) of GdAuGe/graphene/Ge(111). (a) Wide field of view showing a highly ordered film with no boundary between seed and continued growth. (b) Zoom in showing a sharp GdAuGe/graphene/Ge interface.}
    \label{tem}
\end{figure*}

Cross-sectional high angular annular dark field (HAADF) STEM imaging of the seeded sample reveals a well ordered GdAuGe /graphene/Ge interface (Figure \ref{tem}). The GdAuGe /graphene/Ge interface appears atomically sharp, terminated with an Au/Ge plane above the graphene interface. Due to Z-contrast image intensity, heavier elements such as Gd and Au are bright and easily visible whereas lighter carbon in the graphene layer exhibits a relatively weak signal. Just below the graphene, the top layer of the Ge substrate displays brighter contrast than the bulk Ge substrate, indicating possible Gd intercalation. Gd intercalation under graphene has been reported previously, although at slightly higher temperatures of $800\degree$C \cite{link2019introducing}, compared to the $30\degree$ C deposition and $480\degree$C anneal and regrowth we employ here. Finally, we observe that the film is a continuous single crystal between the annealed seed layer and continued growth layer, with no obvious defects induced by the temperature cycling. Our combined diffraction (XRD and RHEED), surface topography (AFM), and electron microscopy support the enhanced structural quality and smoothness for films grown by our low temperature seed approach compared to direct growth at elevated temperature.

Our low temperature seed layer approach appears generalizable to other graphene-covered substrates. As shown in Supplementary Fig. 3, seeded GdAuGe growth on graphene/SiC (0001) produces films with narrower rocking curve width ($\Delta \omega = 9$ arc sec), smoother AFM morphology, and sharper Kiessig fringes compared to films grown directly at high temperature. Additionally, the seeded samples show the expected metallic transport, compared to the hopping transport observed for high temperature growth due to the disconnected island morphology (Supplementary Fig. 4).

\begin{figure}[h]
    \centering
    \includegraphics[width=.5\textwidth]{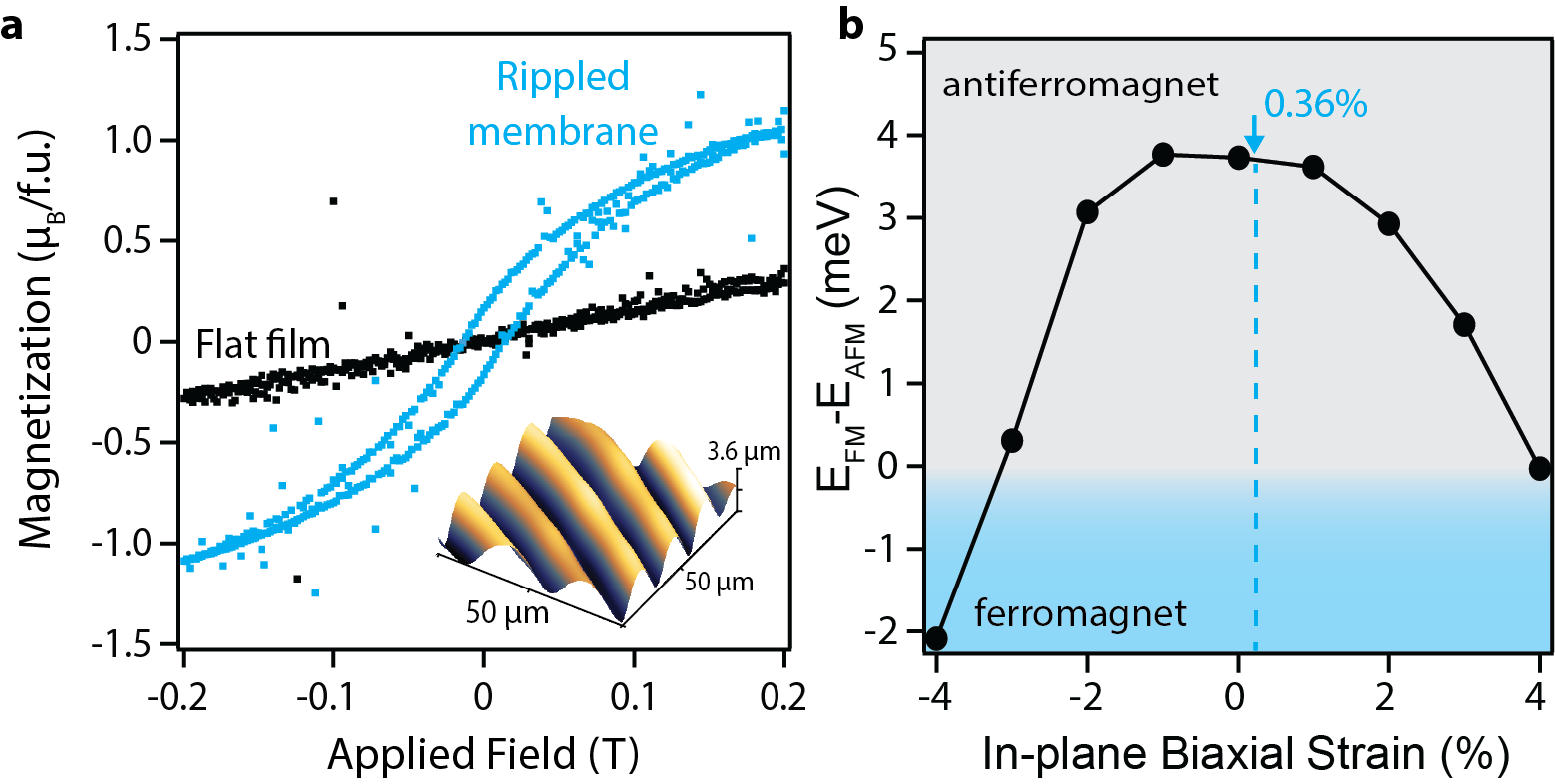}
    \caption{Strain gradient induced ferri/ferromagnetism in a rippled GdAuGe membrane. (a) SQUID magnetometry at 2K with in plane field. Insert: AFM topography of the rippled membrane. (b) DFT calculation of the difference in total energy for ferromagnetic versus A-type antiferromagnetically ordered GdAuGe with moments oriented along the crystallographic $c$ axis.}
    \label{mag}
\end{figure}

Our smooth GdAuGe films grown using the seed approach provide an excellent starting point for membrane exfoliation and manipulation of magnetic properties via extreme strain. To produce membranes with large strain gradients, we adhere a polyurethane handle under tensile strain to the GdAuGe film and exfoliate by cooling the sample in liquid nitrogen. Upon exfoliation, the polyurethane contracts to induce ripples in the GdAuGe membrane. For GdAuGe films grown with a seed layer, this exfoliation produces continuous membranes with well ordered ripples over length scales of several hundred of microns (Fig. \ref{mag}(a) insert and Supplemental Fig. 5). In contrast, exfoliation of samples grown directly at high temperature on graphene/Ge show a rough and disconnected morphology with no ripples, suggesting that the polymer contraction does not produce a well controlled strain pattern in the unseeded samples.

Fig. \ref{mag} compares SQUID magnetometry of a rippled GdAuGe membrane grown using the seed layer, compared with a relaxed GdAuGe film. At the measurement temperature of 2 K, the film shows a weak linear dependence of magnetization on applied field, consistent with the antiferromagnetism observed in bulk crystals \cite{ram2023multiple, kurumaji2024metamagnetism, gibson1996susceptibility}. In contrast, the rippled sample displays a clear hysteresis loop indicative of ferro or ferrimagnetic ordering. The behavior is similar to GdPtSb membranes for which we also observe bending-induced ferri/ferromagnetism \cite{du2021epitaxy}, but with a factor of 5 smaller magnetic saturation than GdPtSb at similar bend radius.

An important distinction is whether the magnetic ordering is induced by homogeneous strain (piezomagnetism and magnetotostriction) or by strain gradients (flexomagnetism). For a smoothly curved membrane with no plastic deformation, the strain along the curved path $s$ is approximately $\varepsilon_s \approx (R+t)/R-1$, where $t$ is the height or depth along the thickness axis and $R$ is the radius of curvature \cite{du2021epitaxy}. Based on the measured thickness of 20 nm, period of 12 $\mu$m, and peak to peak height of 2.6 $\mu$m, we estimate a maximum strain of $\varepsilon_{s} \approx \pm 0.36 \%$ and maximum strain gradient of $\partial \varepsilon_{s} / \partial z \approx \pm 36  \%/\mu m$ at the peaks and valleys. Compared with our density functional theory calculations of antiferromagnetic vs ferromagnetic ordering (Fig \ref{mag}(b) and Supplemental Figure 6), we find that the peak strain in our rippled membrane is an order of magnitude smaller than the calculated homogeneous strain to induce ferromagnetic ordering. This suggests that coupling to strain gradients is more likely responsible for the observed ferro/ferrimagnetic ordering than coupling to homogeneous strain, similar to GdPtSb \cite{du2021epitaxy}. We note, however, that the antiferromagnetic ordering of GdAuGe has not yet been solved \cite{ram2023multiple}, and in our calculation we assume $A$-type ordering with moments pointed along $c$. The particular choice of AFM structure will change the expected strain induced crossover between AFM and FM ordering. Further experiments and theory are required to solve the magnetic structure of GdAuGe and understand its response to strain and strain gradients.

In conclusion, we demonstrated a cold seeded epitaxy approach for growing few nanometer thick, smooth GdAuGe films on graphene-covered Ge and SiC substrates. Seeded films display sharper Bragg reflections, lower mosaicity, and a smoother atomic step and terrace morphology than their counterparts grown directly at high temperature. These films are epitaxial to graphene rather than the underlying Ge or SiC substrate. The interfaces with graphene are atomically sharp. Using a rippled membrane exfoliate from graphene, we demonstrated an antiferromagnetic to ferri/ferromagnetic transition induced by strain gradients. This flexomagnetic response is similar to GdPtSb. Our cold seed strategy provides a path for access to extreme strains in membranes in the ultrathin limit.

\section*{Supporting Information}

Contains the following:

\textbf{Methods} for STEM and XRD.

\textbf{Fig. 1.} Epitaxial orientation of GdAuGe on graphene/Ge (110).

\textbf{Fig. 2.} Composition of GdAuGe films on graphene/Ge by RBS.

\textbf{Fig. 3.} XRD of GdAuGe films on graphene/SiC.

\textbf{Fig. 4.} Electrical transport of GdAuGe films on graphene/SiC.

\textbf{Fig. 5.} Optical microscopy of exfoliated GdAuGe membranes.

\textbf{Fig. 6.} DFT calculated energy difference between FM and AFM configurations for strain along $a$ and $b$ axes.

\section*{Data Availability}

All raw data that appears in this manuscript are posted on the \textcolor{red}{XX repository, doi number YY.}

\section*{Acknowledgments}

We thank Chengye Dong and Joshua Robinson at The Pennsylvania State University Two-Dimensional Crystal Consortium – Materials Innovation Platform (2DCC-MIP) for graphene on SiC samples. 2DCC-MIP is supported by NSF cooperative agreement DMR-2039351. We thank Greg Haugstad at the University of Minnesota Characterization Facility for RBS measurements.

This project was primarily supported by the the Department of Energy Office of Science (DE-SC00203958, Heusler synthesis by Z.L., T.S., T.J., and J.K.K.). Magnetic measurements (Z.L., T.S., and J.K.K.) and electron microscopy (N.H. and P.M.V.) were supported by NSF through the University of Wisconsin Materials Research Science and Engineering Center (DMR-2309000). We acknowledge user facilities also supported by the Wisconsin MRSEC. Transport and preliminary synthesis (Z.L. and J.K.K.) were supported by the Air Force Office of Scientific Research (FA9550-21-0127). Graphene on germanium synthesis and characterization (K.S. and M.S.A.) are supported by the U.S. Department of Energy, Office of Science, Basic Energy Sciences, (DE-SC0016007) and the NSF Graduate Research Fellowship Program (DGE-1747503). DFT calculations (K.T.G. and K.M.R.) were supported by the Office of Naval Research (ONR N00014-21-1-2107) and Wisconsin MRSEC.

\bibliographystyle{apsrev}
\bibliography{ref}

\providecommand{\latin}[1]{#1}
\makeatletter
\providecommand{\doi}
  {\begingroup\let\do\@makeother\dospecials
  \catcode`\{=1 \catcode`\}=2 \doi@aux}
\providecommand{\doi@aux}[1]{\endgroup\texttt{#1}}
\makeatother
\providecommand*\mcitethebibliography{\thebibliography}
\csname @ifundefined\endcsname{endmcitethebibliography}
  {\let\endmcitethebibliography\endthebibliography}{}
\begin{mcitethebibliography}{35}
\providecommand*\natexlab[1]{#1}
\providecommand*\mciteSetBstSublistMode[1]{}
\providecommand*\mciteSetBstMaxWidthForm[2]{}
\providecommand*\mciteBstWouldAddEndPuncttrue
  {\def\EndOfBibitem{\unskip.}}
\providecommand*\mciteBstWouldAddEndPunctfalse
  {\let\EndOfBibitem\relax}
\providecommand*\mciteSetBstMidEndSepPunct[3]{}
\providecommand*\mciteSetBstSublistLabelBeginEnd[3]{}
\providecommand*\EndOfBibitem{}
\mciteSetBstSublistMode{f}
\mciteSetBstMaxWidthForm{subitem}{(\alph{mcitesubitemcount})}
\mciteSetBstSublistLabelBeginEnd
  {\mcitemaxwidthsubitemform\space}
  {\relax}
  {\relax}

\bibitem[Kim \latin{et~al.}(2017)Kim, Cruz, Lee, Alawode, Choi, Song, Johnson,
  Heidelberger, Kong, Choi, \latin{et~al.} others]{kim2017remote}
Kim,~Y.; Cruz,~S.~S.; Lee,~K.; Alawode,~B.~O.; Choi,~C.; Song,~Y.;
  Johnson,~J.~M.; Heidelberger,~C.; Kong,~W.; Choi,~S.; others Remote epitaxy
  through graphene enables two-dimensional material-based layer transfer.
  \emph{Nature} \textbf{2017}, \emph{544}, 340--343\relax
\mciteBstWouldAddEndPuncttrue
\mciteSetBstMidEndSepPunct{\mcitedefaultmidpunct}
{\mcitedefaultendpunct}{\mcitedefaultseppunct}\relax
\EndOfBibitem
\bibitem[Kong \latin{et~al.}(2018)Kong, Li, Qiao, Kim, Lee, Nie, Lee, Osadchy,
  Molnar, Gaskill, \latin{et~al.} others]{kong2018polarity}
Kong,~W.; Li,~H.; Qiao,~K.; Kim,~Y.; Lee,~K.; Nie,~Y.; Lee,~D.; Osadchy,~T.;
  Molnar,~R.~J.; Gaskill,~D.~K.; others Polarity governs atomic interaction
  through two-dimensional materials. \emph{Nature materials} \textbf{2018},
  \emph{17}, 999--1004\relax
\mciteBstWouldAddEndPuncttrue
\mciteSetBstMidEndSepPunct{\mcitedefaultmidpunct}
{\mcitedefaultendpunct}{\mcitedefaultseppunct}\relax
\EndOfBibitem
\bibitem[Koma(1992)]{koma1992van}
Koma,~A. Van der Waals epitaxy—a new epitaxial growth method for a highly
  lattice-mismatched system. \emph{Thin Solid Films} \textbf{1992}, \emph{216},
  72--76\relax
\mciteBstWouldAddEndPuncttrue
\mciteSetBstMidEndSepPunct{\mcitedefaultmidpunct}
{\mcitedefaultendpunct}{\mcitedefaultseppunct}\relax
\EndOfBibitem
\bibitem[Shi \latin{et~al.}(2012)Shi, Zhou, Lu, Fang, Lee, Hsu, Kim, Kim, Yang,
  Li, \latin{et~al.} others]{shi2012van}
Shi,~Y.; Zhou,~W.; Lu,~A.-Y.; Fang,~W.; Lee,~Y.-H.; Hsu,~A.~L.; Kim,~S.~M.;
  Kim,~K.~K.; Yang,~H.~Y.; Li,~L.-J.; others van der Waals epitaxy of MoS2
  layers using graphene as growth templates. \emph{Nano letters} \textbf{2012},
  \emph{12}, 2784--2791\relax
\mciteBstWouldAddEndPuncttrue
\mciteSetBstMidEndSepPunct{\mcitedefaultmidpunct}
{\mcitedefaultendpunct}{\mcitedefaultseppunct}\relax
\EndOfBibitem
\bibitem[Kim \latin{et~al.}(2014)Kim, Bayram, Park, Cheng, Dimitrakopoulos,
  Ott, Reuter, Bedell, and Sadana]{kim2014principle}
Kim,~J.; Bayram,~C.; Park,~H.; Cheng,~C.-W.; Dimitrakopoulos,~C.; Ott,~J.~A.;
  Reuter,~K.~B.; Bedell,~S.~W.; Sadana,~D.~K. Principle of direct van der Waals
  epitaxy of single-crystalline films on epitaxial graphene. \emph{Nature
  communications} \textbf{2014}, \emph{5}, 4836\relax
\mciteBstWouldAddEndPuncttrue
\mciteSetBstMidEndSepPunct{\mcitedefaultmidpunct}
{\mcitedefaultendpunct}{\mcitedefaultseppunct}\relax
\EndOfBibitem
\bibitem[Du \latin{et~al.}(2022)Du, Jung, Manzo, LaDuca, Zheng, Su, Saraswat,
  McChesney, Arnold, and Kawasaki]{du2022controlling}
Du,~D.; Jung,~T.; Manzo,~S.; LaDuca,~Z.; Zheng,~X.; Su,~K.; Saraswat,~V.;
  McChesney,~J.; Arnold,~M.~S.; Kawasaki,~J.~K. Controlling the balance between
  remote, pinhole, and van der Waals epitaxy of Heusler films on
  graphene/sapphire. \emph{Nano Letters} \textbf{2022}, \emph{22},
  8647--8653\relax
\mciteBstWouldAddEndPuncttrue
\mciteSetBstMidEndSepPunct{\mcitedefaultmidpunct}
{\mcitedefaultendpunct}{\mcitedefaultseppunct}\relax
\EndOfBibitem
\bibitem[Hong \latin{et~al.}(2020)Hong, Gu, Verma, Harbola, Wang, Lu,
  Vailionis, Hikita, Pentcheva, Rondinelli, \latin{et~al.}
  others]{hong2020extreme}
Hong,~S.~S.; Gu,~M.; Verma,~M.; Harbola,~V.; Wang,~B.~Y.; Lu,~D.;
  Vailionis,~A.; Hikita,~Y.; Pentcheva,~R.; Rondinelli,~J.~M.; others Extreme
  tensile strain states in La0. 7Ca0. 3MnO3 membranes. \emph{Science}
  \textbf{2020}, \emph{368}, 71--76\relax
\mciteBstWouldAddEndPuncttrue
\mciteSetBstMidEndSepPunct{\mcitedefaultmidpunct}
{\mcitedefaultendpunct}{\mcitedefaultseppunct}\relax
\EndOfBibitem
\bibitem[Harbola \latin{et~al.}(2021)Harbola, Crossley, Hong, Lu, Birkholzer,
  Hikita, and Hwang]{harbola2021strain}
Harbola,~V.; Crossley,~S.; Hong,~S.~S.; Lu,~D.; Birkholzer,~Y.~A.; Hikita,~Y.;
  Hwang,~H.~Y. Strain gradient elasticity in SrTiO3 membranes: bending versus
  stretching. \emph{Nano letters} \textbf{2021}, \emph{21}, 2470--2475\relax
\mciteBstWouldAddEndPuncttrue
\mciteSetBstMidEndSepPunct{\mcitedefaultmidpunct}
{\mcitedefaultendpunct}{\mcitedefaultseppunct}\relax
\EndOfBibitem
\bibitem[Du \latin{et~al.}(2021)Du, Manzo, Zhang, Saraswat, Genser, Rabe,
  Voyles, Arnold, and Kawasaki]{du2021epitaxy}
Du,~D.; Manzo,~S.; Zhang,~C.; Saraswat,~V.; Genser,~K.~T.; Rabe,~K.~M.;
  Voyles,~P.~M.; Arnold,~M.~S.; Kawasaki,~J.~K. Epitaxy, exfoliation, and
  strain-induced magnetism in rippled Heusler membranes. \emph{Nature
  communications} \textbf{2021}, \emph{12}, 1--7\relax
\mciteBstWouldAddEndPuncttrue
\mciteSetBstMidEndSepPunct{\mcitedefaultmidpunct}
{\mcitedefaultendpunct}{\mcitedefaultseppunct}\relax
\EndOfBibitem
\bibitem[Dong \latin{et~al.}(2019)Dong, Li, Yao, Zhou, Zhang, Han, Luo, Yao,
  Peng, Hu, \latin{et~al.} others]{dong2019super}
Dong,~G.; Li,~S.; Yao,~M.; Zhou,~Z.; Zhang,~Y.-Q.; Han,~X.; Luo,~Z.; Yao,~J.;
  Peng,~B.; Hu,~Z.; others Super-elastic ferroelectric single-crystal membrane
  with continuous electric dipole rotation. \emph{Science} \textbf{2019},
  \emph{366}, 475--479\relax
\mciteBstWouldAddEndPuncttrue
\mciteSetBstMidEndSepPunct{\mcitedefaultmidpunct}
{\mcitedefaultendpunct}{\mcitedefaultseppunct}\relax
\EndOfBibitem
\bibitem[Lukashev and Sabirianov(2010)Lukashev, and
  Sabirianov]{lukashev2010flexomagnetic}
Lukashev,~P.; Sabirianov,~R.~F. Flexomagnetic effect in frustrated triangular
  magnetic structures. \emph{Physical Review B} \textbf{2010}, \emph{82},
  094417\relax
\mciteBstWouldAddEndPuncttrue
\mciteSetBstMidEndSepPunct{\mcitedefaultmidpunct}
{\mcitedefaultendpunct}{\mcitedefaultseppunct}\relax
\EndOfBibitem
\bibitem[Zhang \latin{et~al.}(2021)Zhang, Liu, Dong, Wu, Zhang, Wang, Lu,
  R{\"u}ckriegel, Wang, Duine, \latin{et~al.} others]{zhang2021strain}
Zhang,~Y.; Liu,~J.; Dong,~Y.; Wu,~S.; Zhang,~J.; Wang,~J.; Lu,~J.;
  R{\"u}ckriegel,~A.; Wang,~H.; Duine,~R.; others Strain-driven
  Dzyaloshinskii-Moriya interaction for room-temperature magnetic skyrmions.
  \emph{Physical Review Letters} \textbf{2021}, \emph{127}, 117204\relax
\mciteBstWouldAddEndPuncttrue
\mciteSetBstMidEndSepPunct{\mcitedefaultmidpunct}
{\mcitedefaultendpunct}{\mcitedefaultseppunct}\relax
\EndOfBibitem
\bibitem[Makushko \latin{et~al.}(2022)Makushko, Kosub, Pylypovskyi, Hedrich,
  Li, Pashkin, Avdoshenko, H{\"u}bner, Ganss, Wolf, \latin{et~al.}
  others]{makushko2022flexomagnetism}
Makushko,~P.; Kosub,~T.; Pylypovskyi,~O.~V.; Hedrich,~N.; Li,~J.; Pashkin,~A.;
  Avdoshenko,~S.; H{\"u}bner,~R.; Ganss,~F.; Wolf,~D.; others Flexomagnetism
  and vertically graded N{\'e}el temperature of antiferromagnetic Cr2O3 thin
  films. \emph{Nature Communications} \textbf{2022}, \emph{13}, 6745\relax
\mciteBstWouldAddEndPuncttrue
\mciteSetBstMidEndSepPunct{\mcitedefaultmidpunct}
{\mcitedefaultendpunct}{\mcitedefaultseppunct}\relax
\EndOfBibitem
\bibitem[Du \latin{et~al.}(2023)Du, Hu, and Kawasaki]{du2023strain}
Du,~D.; Hu,~J.; Kawasaki,~J.~K. Strain and strain gradient engineering in
  membranes of quantum materials. \emph{Applied Physics Letters} \textbf{2023},
  \emph{122}\relax
\mciteBstWouldAddEndPuncttrue
\mciteSetBstMidEndSepPunct{\mcitedefaultmidpunct}
{\mcitedefaultendpunct}{\mcitedefaultseppunct}\relax
\EndOfBibitem
\bibitem[Edstr{\"o}m \latin{et~al.}(2022)Edstr{\"o}m, Amoroso, Picozzi, Barone,
  and Stengel]{edstrom2022curved}
Edstr{\"o}m,~A.; Amoroso,~D.; Picozzi,~S.; Barone,~P.; Stengel,~M. Curved
  magnetism in CrI 3. \emph{Physical review letters} \textbf{2022}, \emph{128},
  177202\relax
\mciteBstWouldAddEndPuncttrue
\mciteSetBstMidEndSepPunct{\mcitedefaultmidpunct}
{\mcitedefaultendpunct}{\mcitedefaultseppunct}\relax
\EndOfBibitem
\bibitem[Ga \latin{et~al.}(2022)Ga, Cui, Liang, Yu, Zhu, Wang, and
  Yang]{ga2022dzyaloshinskii}
Ga,~Y.; Cui,~Q.; Liang,~J.; Yu,~D.; Zhu,~Y.; Wang,~L.; Yang,~H.
  Dzyaloshinskii-Moriya interaction and magnetic skyrmions induced by
  curvature. \emph{Physical Review B} \textbf{2022}, \emph{106}, 054426\relax
\mciteBstWouldAddEndPuncttrue
\mciteSetBstMidEndSepPunct{\mcitedefaultmidpunct}
{\mcitedefaultendpunct}{\mcitedefaultseppunct}\relax
\EndOfBibitem
\bibitem[Qiao \latin{et~al.}(2021)Qiao, Liu, Kim, Molnar, Osadchy, Li, Sun, Li,
  Myers-Ward, Lee, \latin{et~al.} others]{qiao2021graphene}
Qiao,~K.; Liu,~Y.; Kim,~C.; Molnar,~R.~J.; Osadchy,~T.; Li,~W.; Sun,~X.;
  Li,~H.; Myers-Ward,~R.~L.; Lee,~D.; others Graphene buffer layer on SiC as a
  release layer for high-quality freestanding semiconductor membranes.
  \emph{Nano letters} \textbf{2021}, \emph{21}, 4013--4020\relax
\mciteBstWouldAddEndPuncttrue
\mciteSetBstMidEndSepPunct{\mcitedefaultmidpunct}
{\mcitedefaultendpunct}{\mcitedefaultseppunct}\relax
\EndOfBibitem
\bibitem[Chang \latin{et~al.}(2023)Chang, Kim, Park, Choi, Kim, Jeong, Barone,
  Parker, Lee, Zhang, \latin{et~al.} others]{chang2023remote}
Chang,~C.~S.; Kim,~K.~S.; Park,~B.-I.; Choi,~J.; Kim,~H.; Jeong,~J.;
  Barone,~M.; Parker,~N.; Lee,~S.; Zhang,~X.; others Remote epitaxial
  interaction through graphene. \emph{Science Advances} \textbf{2023},
  \emph{9}, eadj5379\relax
\mciteBstWouldAddEndPuncttrue
\mciteSetBstMidEndSepPunct{\mcitedefaultmidpunct}
{\mcitedefaultendpunct}{\mcitedefaultseppunct}\relax
\EndOfBibitem
\bibitem[Alaskar \latin{et~al.}(2014)Alaskar, Arafin, Wickramaratne, Zurbuchen,
  He, McKay, Lin, Goorsky, Lake, and Wang]{alaskar2014towards}
Alaskar,~Y.; Arafin,~S.; Wickramaratne,~D.; Zurbuchen,~M.~A.; He,~L.;
  McKay,~J.; Lin,~Q.; Goorsky,~M.~S.; Lake,~R.~K.; Wang,~K.~L. Towards van der
  Waals epitaxial growth of GaAs on Si using a graphene buffer layer.
  \emph{Advanced Functional Materials} \textbf{2014}, \emph{24},
  6629--6638\relax
\mciteBstWouldAddEndPuncttrue
\mciteSetBstMidEndSepPunct{\mcitedefaultmidpunct}
{\mcitedefaultendpunct}{\mcitedefaultseppunct}\relax
\EndOfBibitem
\bibitem[Zulqurnain \latin{et~al.}(2022)Zulqurnain, Burton, Al-Hada, Goff,
  Hofmann, and Hirst]{zulqurnain2022defect}
Zulqurnain,~M.; Burton,~O.~J.; Al-Hada,~M.; Goff,~L.~E.; Hofmann,~S.;
  Hirst,~L.~C. Defect seeded remote epitaxy of GaAs films on graphene.
  \emph{Nanotechnology} \textbf{2022}, \emph{33}, 485603\relax
\mciteBstWouldAddEndPuncttrue
\mciteSetBstMidEndSepPunct{\mcitedefaultmidpunct}
{\mcitedefaultendpunct}{\mcitedefaultseppunct}\relax
\EndOfBibitem
\bibitem[Manzo \latin{et~al.}(2022)Manzo, Strohbeen, Lim, Saraswat, Du, Xu,
  Pokharel, Mawst, Arnold, and Kawasaki]{manzo2022pinhole}
Manzo,~S.; Strohbeen,~P.~J.; Lim,~Z.~H.; Saraswat,~V.; Du,~D.; Xu,~S.;
  Pokharel,~N.; Mawst,~L.~J.; Arnold,~M.~S.; Kawasaki,~J.~K. Pinhole-seeded
  lateral epitaxy and exfoliation of GaSb films on graphene-terminated
  surfaces. \emph{Nature communications} \textbf{2022}, \emph{13}, 1--9\relax
\mciteBstWouldAddEndPuncttrue
\mciteSetBstMidEndSepPunct{\mcitedefaultmidpunct}
{\mcitedefaultendpunct}{\mcitedefaultseppunct}\relax
\EndOfBibitem
\bibitem[Hong \latin{et~al.}(2012)Hong, Lee, Wu, Ruoff, and Fukui]{hong2012van}
Hong,~Y.~J.; Lee,~W.~H.; Wu,~Y.; Ruoff,~R.~S.; Fukui,~T. van der Waals epitaxy
  of InAs nanowires vertically aligned on single-layer graphene. \emph{Nano
  letters} \textbf{2012}, \emph{12}, 1431--1436\relax
\mciteBstWouldAddEndPuncttrue
\mciteSetBstMidEndSepPunct{\mcitedefaultmidpunct}
{\mcitedefaultendpunct}{\mcitedefaultseppunct}\relax
\EndOfBibitem
\bibitem[Liu \latin{et~al.}(2013)Liu, Wang, Hupalo, Lin, Ho, and
  Tringides]{liu2013metals}
Liu,~X.; Wang,~C.-Z.; Hupalo,~M.; Lin,~H.-Q.; Ho,~K.-M.; Tringides,~M.~C.
  Metals on graphene: interactions, growth morphology, and thermal stability.
  \emph{Crystals} \textbf{2013}, \emph{3}, 79--111\relax
\mciteBstWouldAddEndPuncttrue
\mciteSetBstMidEndSepPunct{\mcitedefaultmidpunct}
{\mcitedefaultendpunct}{\mcitedefaultseppunct}\relax
\EndOfBibitem
\bibitem[LaDuca \latin{et~al.}(2023)LaDuca, Su, Manzo, Arnold, and
  Kawasaki]{laduca2023control}
LaDuca,~Z.; Su,~K.; Manzo,~S.; Arnold,~M.~S.; Kawasaki,~J.~K. Control of
  ternary alloy composition during remote epitaxy on graphene. \emph{Physical
  Review Materials} \textbf{2023}, \emph{7}, 083401\relax
\mciteBstWouldAddEndPuncttrue
\mciteSetBstMidEndSepPunct{\mcitedefaultmidpunct}
{\mcitedefaultendpunct}{\mcitedefaultseppunct}\relax
\EndOfBibitem
\bibitem[Kiraly \latin{et~al.}(2015)Kiraly, Jacobberger, Mannix, Campbell,
  Bedzyk, Arnold, Hersam, and Guisinger]{kiraly2015electronic}
Kiraly,~B.; Jacobberger,~R.~M.; Mannix,~A.~J.; Campbell,~G.~P.; Bedzyk,~M.~J.;
  Arnold,~M.~S.; Hersam,~M.~C.; Guisinger,~N.~P. Electronic and mechanical
  properties of graphene--germanium interfaces grown by chemical vapor
  deposition. \emph{Nano letters} \textbf{2015}, \emph{15}, 7414--7420\relax
\mciteBstWouldAddEndPuncttrue
\mciteSetBstMidEndSepPunct{\mcitedefaultmidpunct}
{\mcitedefaultendpunct}{\mcitedefaultseppunct}\relax
\EndOfBibitem
\bibitem[Badokas \latin{et~al.}(2021)Badokas, Kadys, Mickevi{\v{c}}ius,
  Ignatjev, Skapas, Stanionyt{\.e}, Radiunas, Ju{\v{s}}ka, and
  Malinauskas]{badokas2021remote}
Badokas,~K.; Kadys,~A.; Mickevi{\v{c}}ius,~J.; Ignatjev,~I.; Skapas,~M.;
  Stanionyt{\.e},~S.; Radiunas,~E.; Ju{\v{s}}ka,~G.; Malinauskas,~T. Remote
  epitaxy of GaN via graphene on GaN/sapphire templates. \emph{Journal of
  Physics D: Applied Physics} \textbf{2021}, \emph{54}, 205103\relax
\mciteBstWouldAddEndPuncttrue
\mciteSetBstMidEndSepPunct{\mcitedefaultmidpunct}
{\mcitedefaultendpunct}{\mcitedefaultseppunct}\relax
\EndOfBibitem
\bibitem[Yoon \latin{et~al.}(2022)Yoon, Truttmann, Liu, Matthews, Choo, Su,
  Saraswat, Manzo, Arnold, Bowden, \latin{et~al.} others]{yoon2022free}
Yoon,~H.; Truttmann,~T.~K.; Liu,~F.; Matthews,~B.~E.; Choo,~S.; Su,~Q.;
  Saraswat,~V.; Manzo,~S.; Arnold,~M.~S.; Bowden,~M.~E.; others Free-Standing
  Epitaxial SrTiO $ \_3 $ Nanomembranes via Remote Epitaxy using Hybrid
  Molecular Beam Epitaxy. \emph{arXiv preprint arXiv:2206.09094} \textbf{2022},
  \relax
\mciteBstWouldAddEndPunctfalse
\mciteSetBstMidEndSepPunct{\mcitedefaultmidpunct}
{}{\mcitedefaultseppunct}\relax
\EndOfBibitem
\bibitem[Guo \latin{et~al.}(2019)Guo, Sun, Jiang, Wang, Chen, Yin, Qi, Gao,
  Zhang, Lu, \latin{et~al.} others]{guo2019reconfigurable}
Guo,~Y.; Sun,~X.; Jiang,~J.; Wang,~B.; Chen,~X.; Yin,~X.; Qi,~W.; Gao,~L.;
  Zhang,~L.; Lu,~Z.; others A reconfigurable remotely epitaxial VO2 electrical
  heterostructure. \emph{Nano letters} \textbf{2019}, \emph{20}, 33--42\relax
\mciteBstWouldAddEndPuncttrue
\mciteSetBstMidEndSepPunct{\mcitedefaultmidpunct}
{\mcitedefaultendpunct}{\mcitedefaultseppunct}\relax
\EndOfBibitem
\bibitem[Wang \latin{et~al.}(2022)Wang, Wang, Wang, He, Huang, Pan, Zhu, Wang,
  and Ye]{wang2022improved}
Wang,~N.; Wang,~P.; Wang,~F.; He,~H.; Huang,~J.; Pan,~X.; Zhu,~G.; Wang,~J.;
  Ye,~Z. Improved epitaxy of ZnO films by regulating the layers of graphene.
  \emph{Applied Surface Science} \textbf{2022}, \emph{585}, 152709\relax
\mciteBstWouldAddEndPuncttrue
\mciteSetBstMidEndSepPunct{\mcitedefaultmidpunct}
{\mcitedefaultendpunct}{\mcitedefaultseppunct}\relax
\EndOfBibitem
\bibitem[Dai \latin{et~al.}(2022)Dai, Zhao, Li, Chen, Zhai, Xue, Di, Feng, Sun,
  Luo, \latin{et~al.} others]{dai2022highly}
Dai,~L.; Zhao,~J.; Li,~J.; Chen,~B.; Zhai,~S.; Xue,~Z.; Di,~Z.; Feng,~B.;
  Sun,~Y.; Luo,~Y.; others Highly heterogeneous epitaxy of flexoelectric
  BaTiO3-$\delta$ membrane on Ge. \emph{Nature Communications} \textbf{2022},
  \emph{13}, 1--10\relax
\mciteBstWouldAddEndPuncttrue
\mciteSetBstMidEndSepPunct{\mcitedefaultmidpunct}
{\mcitedefaultendpunct}{\mcitedefaultseppunct}\relax
\EndOfBibitem
\bibitem[Link \latin{et~al.}(2019)Link, Forti, St{\"o}hr, K{\"u}ster,
  R{\"o}sner, Hirschmeier, Chen, Avila, Asensio, Zakharov, \latin{et~al.}
  others]{link2019introducing}
Link,~S.; Forti,~S.; St{\"o}hr,~A.; K{\"u}ster,~K.; R{\"o}sner,~M.;
  Hirschmeier,~D.; Chen,~C.; Avila,~J.; Asensio,~M.; Zakharov,~A.; others
  Introducing strong correlation effects into graphene by gadolinium
  intercalation. \emph{Physical Review B} \textbf{2019}, \emph{100},
  121407\relax
\mciteBstWouldAddEndPuncttrue
\mciteSetBstMidEndSepPunct{\mcitedefaultmidpunct}
{\mcitedefaultendpunct}{\mcitedefaultseppunct}\relax
\EndOfBibitem
\bibitem[Ram \latin{et~al.}(2023)Ram, Singh, Hooda, Singh, Kanchana,
  Kaczorowski, and Hossain]{ram2023multiple}
Ram,~D.; Singh,~J.; Hooda,~M.; Singh,~K.; Kanchana,~V.; Kaczorowski,~D.;
  Hossain,~Z. Multiple magnetic transitions, metamagnetism, and large
  magnetoresistance in GdAuGe single crystals. \emph{Physical Review B}
  \textbf{2023}, \emph{108}, 235107\relax
\mciteBstWouldAddEndPuncttrue
\mciteSetBstMidEndSepPunct{\mcitedefaultmidpunct}
{\mcitedefaultendpunct}{\mcitedefaultseppunct}\relax
\EndOfBibitem
\bibitem[Kurumaji \latin{et~al.}(2024)Kurumaji, Gen, Kitou, and
  Arima]{kurumaji2024metamagnetism}
Kurumaji,~T.; Gen,~M.; Kitou,~S.; Arima,~T.-h. Metamagnetism and anomalous
  magnetotransport properties in rare-earth-based polar semimetals $ R $ AuGe
  ($ R= $ Dy, Ho, and Gd). \emph{arXiv preprint arXiv:2405.00628}
  \textbf{2024}, \relax
\mciteBstWouldAddEndPunctfalse
\mciteSetBstMidEndSepPunct{\mcitedefaultmidpunct}
{}{\mcitedefaultseppunct}\relax
\EndOfBibitem
\bibitem[Gibson \latin{et~al.}(1996)Gibson, Schnelle, P{\"o}ttgen, Bartkowski,
  and Kremer]{gibson1996susceptibility}
Gibson,~B.; Schnelle,~W.; P{\"o}ttgen,~R.; Bartkowski,~K.; Kremer,~R.
  Susceptibility, specific heat, and transport properties of CeAuGe and GdAuGe.
  \emph{Czechoslovak Journal of Physics} \textbf{1996}, \emph{46},
  2573--2574\relax
\mciteBstWouldAddEndPuncttrue
\mciteSetBstMidEndSepPunct{\mcitedefaultmidpunct}
{\mcitedefaultendpunct}{\mcitedefaultseppunct}\relax
\EndOfBibitem
\end{mcitethebibliography}

\end{document}